\documentclass[twocolumn,showpacs,preprintnumbers]{revtex4}
\usepackage{amsmath}
\usepackage{amssymb}
\usepackage{graphicx}% Include figure files\usepackage{dcolumn}
\usepackage{bm}% bold math
\begin{document}
%************************************************************************

\title{The Dynamical Dipole Mode in Fusion Reactions with Exotic Nuclear Beams}

\author{V.Baran$^{a}$, C.Rizzo $^{b,c}$, M. Colonna$^{b,c}$, 
M.Di Toro$^{b,c,*}$, D. Pierroutsakou $^{d}$}

\affiliation{
        $^a$NIPNE-HH, Bucharest and Bucharest University, Romania\\
        $^b$LNS-INFN, I-95123, Catania, Italy\\
	$^c$Physics and Astronomy Dept. University of Catania, Italy\\
        $^d$INFN, Sezione di Napoli, Italy\\
        $^*$ email: ditoro@lns.infn.it}

\begin{abstract}
% Text of abstract
We report the properties of the prompt dipole radiation, produced via a 
collective bremsstrahlung mechanism, in fusion reactions 
with exotic beams. We show that the gamma yield is sensitive to the density 
dependence of the symmetry energy below/around saturation. Moreover we 
find that 
the angular distribution of the emitted photons 
from such fast collective mode can represent a sensitive probe of its 
excitation mechanism
and of fusion dynamics in the entrance channel.
\end{abstract}

%\keyword{
%Dynamical Dipole, Symmetry energy, Fusion reactions, Photon Angular
%distributions} 
\pacs{25.60.Pj;25.70.Jj;24.30.Cz;21.30.Fe}
\maketitle
\date{\today}
% main text
%\section{}
%\label{}

Production of exotic nuclei has opened the way to explore, in 
laboratory conditions, 
new aspects of nuclear structure and dynamics
up to extreme ratios of neutron (N) to proton numbers (Z). 
An important issue addressed
is the density dependence of symmetry energy term in the nuclear
Equation of State, of interest also for the properties of
astrophysical objects  \cite{bao01,ste05,bar05a}. 
By employing heavy ions collisions at
appropriate beam energy and centrality the
isospin dynamics at different densities 
of nuclear matter can be investigated
\cite{bar05a,xu00,tsa04,bar04,fil05}.
   
Here we discuss isospin effects in dissipative collisions at low energies, 
between $5$ and $20$ MeV/n, were unstable ion beams with large asymmetry
will be soon available. 
The starting point is that in this energy range, for dissipative reactions
between nuclei with different $N/Z$ ratios, the charge equilibration process in
the entrance channel
has a collective character resembling
a large amplitude Giant Dipole Resonance (GDR). 
Several
microscopic transport simulations like
semiclassical Boltzmann-Nordheim-Vlasov (BNV)
\cite{cho93,bar96}, Time-Dependent Hartree-Fock (TDHF)
\cite{bon81,umar85,sim01,umar07}  or Constrained Molecular Dynamics (CoMD) 
\cite{pap03} have confirmed this scenario.
Several experiments based on a comparison between two reactions with 
different entrance channel charge 
asymmetry, leading to the same final products, have confirmed 
the presence of this effect, see 
\cite{fli}, \cite{pie05} and refs. therein.

The gamma yield resulting from the decay of such pre-equilibrium
isovector mode can encode information about the early stage of the 
reaction \cite{cho93,bar96,sim01,bar01b,bar01}. 
This collective response is appearing in the 
lower density intermediate neck region, while the system is still in a 
highly deformed dinuclear configuration.
It is therefore of interest to look at the influence of density 
dependence of symmetry
energy below saturation upon the excitation and dynamics of 
the prompt dipole mode.
The corresponding emission rates
can be evaluated, 
through a ''bremsstrahlung'' mechanism, in a consistent transport 
approach to the 
rection dynamics, which can account for
the whole contribution along
the dissipative nonequilibrium path, in fusion or deep-inelastic processes
\cite{bar01}. In this way the data can be directly used to probe
the isovector part of the in medium effective 
interaction below saturation density.

Here we discuss also the expected emission anisotropy of such prompt 
dipole radiation, due to the neutron/proton oscillation along a definite 
symmetry axis.
This is of interest for planning new experiments and for the relation to 
the lifetime
of such transient collective mode.

\vskip 0.5cm
\noindent

{\it Exotic Beams}

In the following we shall study the features of the pre-equilibrium dipole
considering the reaction $^{132}Sn+^{58}Ni$ (``132'' system) at $10MeV/A$,
 as referred to the same reaction induced by a $^{124}Sn$ beam 
(``124'' system).
From simple arguments a larger initial dipole moment will trigger higher 
amplitude isovector oscillations
increasing the chance of a clear experimental observation.
During the reaction dynamics the dipole moment is given by 
$D(t)=\frac{NZ}{A}X(t)$, where  $A=N+Z$, and $N=N_1+N_2$, $Z=Z_1+Z_2$, are
the total number of participating nucleons, while $X(t)$ is the distance 
between
the centers of mass of protons and neutrons. We note that the initial dipole
($t=0$: touching configuration)   attains a value around
$45 fm$ for the exotic $^{132}Sn$ beam, to be compared to the smaller 
value $33 fm$ 
for the stable ``124'' system, which can be considered as a reference 
partner in an experimental
comparison. 

We have employed a mean field transport approach,
based on the BNV equation, which properly describes  the
selfconsistent couplings between various degrees of freedom.  
The potential part of the symmetry energy, %$C(\rho)$, 
$E_{sym}/A(pot)$:
\begin{equation}
\frac{E_{sym}}{A}=\frac{E_{sym}}{A}(kin)+\frac{E_{sym}}{A}(pot)\equiv 
\frac{\epsilon_F}{3} + \frac{C(\rho)}{2\rho_0}\rho
\end{equation}
 is tested by employing two different density
parametrizations, Isovector Equation of State (Iso-EoS)
of the  mean field:
i) $\frac{C(\rho)}{\rho_0}=482-1638 \rho$, $(MeV fm^{3})$, for ``Asysoft'' 
EoS,  where ${E_{sym}/{A}}(pot)$
has a weak
density dependence close to the saturation, with an almost flat behavior below
 $\rho_0$; ii) a constant coefficient, $C=32 MeV$, for the ``Asystiff'' EoS 
choice, where the interaction part of the symmetry term displays
a linear density dependence. As shown in details in refs. \cite{bar05a,bao08}
these choices represent two classes of widely used effective interactions,
that still require some confirmation from new independent observables.  
The isoscalar section of the EoS is the same in both cases,
corresponding to  a compressibility around $220MeV$.
In the numerical simulations
a test particle approach with 200 gaussian test particles per nucleon has
been employed. In this way we get
 a good description of the phase space occupation, essential for the low 
energy reaction dynamics.
In the collision integral
in medium nucleon-nucleon (N-N) cross sections are considered, \cite{li93}.
We perform calculations for three impact parameters: b = 0,2,4 fm, to cover 
the region where fusion is mostly observed. 
In order to reduce the numerical noise we run twenty events for each set 
of macroscopic initial conditions
and the displayed quantities are the averages over this ensemble.

In Figure 1 we report some global information concerning the dipole mode
in entrance channel. 
\begin{figure}
\begin{center}
\includegraphics*[scale=0.33]{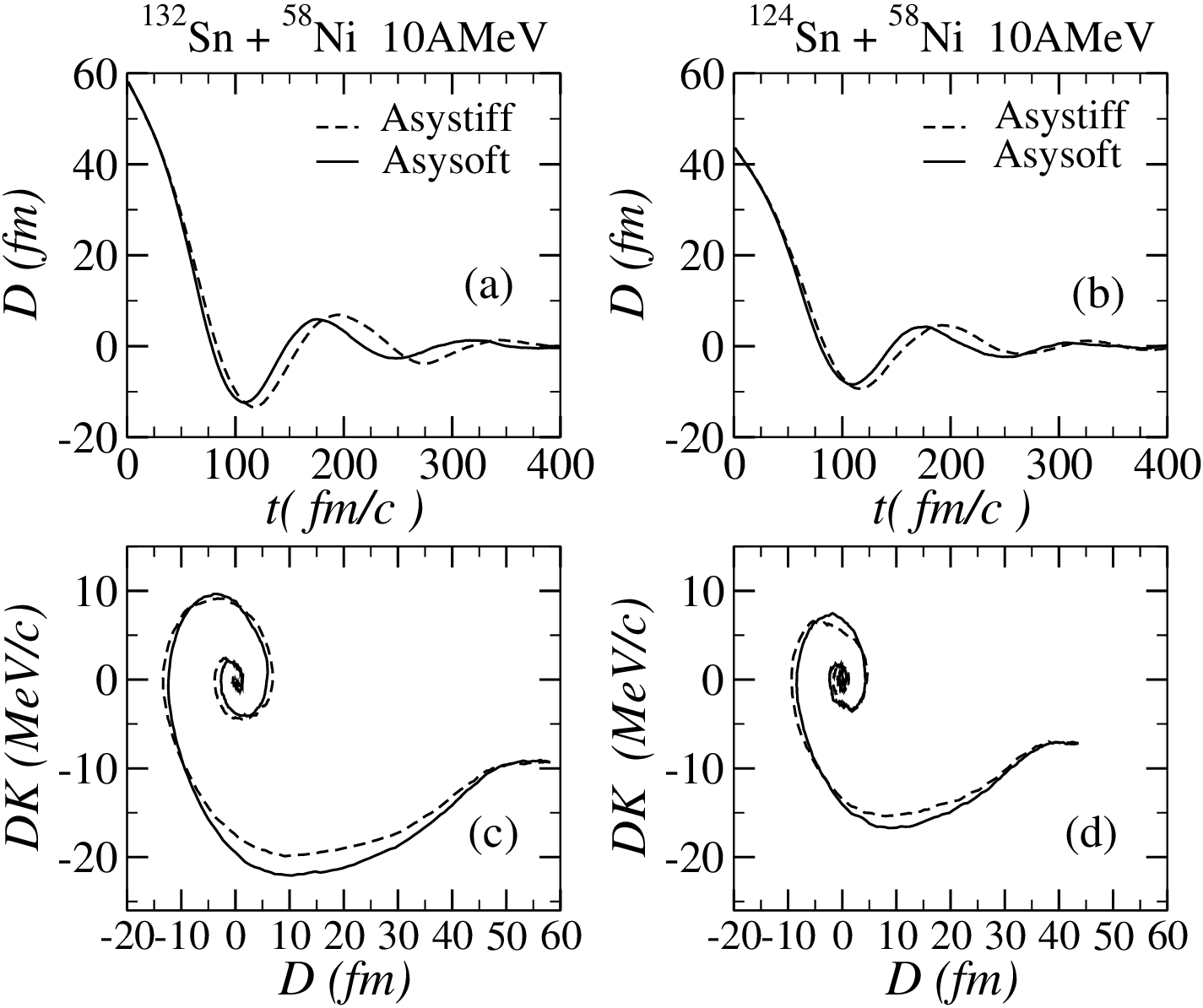}
\end{center}
\caption{Dipole Dynamics at 10AMeV, $b=4fm$ centrality. 
Exotic ``132'' system: (a) Time evolution of dipole moment 
D(t) in real 
space; (c) Dipole phase-space correlation (see text).
Panels (b) and (d): same as before for the stable ``124'' system.
Solid lines correspond to Asysoft EoS, the dashed to Asystiff EoS.}
\label{dip}
\end{figure}
The time evolution of the dipole moment $D(t)$
for the ``132'' system at $b=4fm$ is represented in Fig.1 (a).
A similar behavior is seen at $b=0,2~fm$.
We notice the large amplitude of the first oscillation
and the delayed dynamics for the Asystiff EOS related to the weaker 
isovector
restoring force. We can also evaluate the quantity 
$DK(t)=(\frac{P_p}{Z}-\frac{P_n}{N})$,
the canonically conjugate momentum of the $X(t)$ coordinate.
The phase space correlation (spiraling) between $D(t)$  
and $DK(t)$ 
is reported in Fig.1 (c). 
It nicely points out a collective behavior which initiates very early,
with a dipole moment close to the touching configuration value
reported above.
This can be explained by the fast formation of a well developed neck mean field
which sustains the collective
dipole oscillation in spite of the dinuclear configuration
with a central zone still at densities below the saturation value.

The role of a large charge asymmetry between 
the two 
colliding nuclei can be seen from
Fig.1,(b,d) panels, where we show the analogous dipole phase space 
trajectories for the stable  
$^{124}Sn+^{58}Ni$ system at the same value of impact parameter and energy. 
A clear 
reduction of the 
collective behavior is evidenced. 

In order to appreciate if these differences can be observed experimentally
we estimate the gamma yield in a bremsstrahlung
approach \cite{bar01,class08}: 

\begin{equation}
\frac{dP}{dE_\gamma}=\frac{2e^2}{3 \pi \hbar c^3 E_\gamma}
\mid D''(\omega)\mid ^2,
\label{emission}
\end{equation}
where $E_\gamma=\hbar \omega$ is the photon energy and
$\mid D''(\omega)\mid^2$ is the Fourier transform of the 
dipole ``acceleration'' $
 D''(\omega)=\int_{t_0}^{t_{max}} D''(t) e^{i \omega t} dt.$
For each event $t_0$ represents the onset-time of the collective dipole 
response (phase-space spiraling) and $t_{max}$ the ``damping time''.
i.e. the time step corresponding to an almost flat D(t) behavior.

\begin{figure}
\begin{center}
\includegraphics*[scale=0.33]{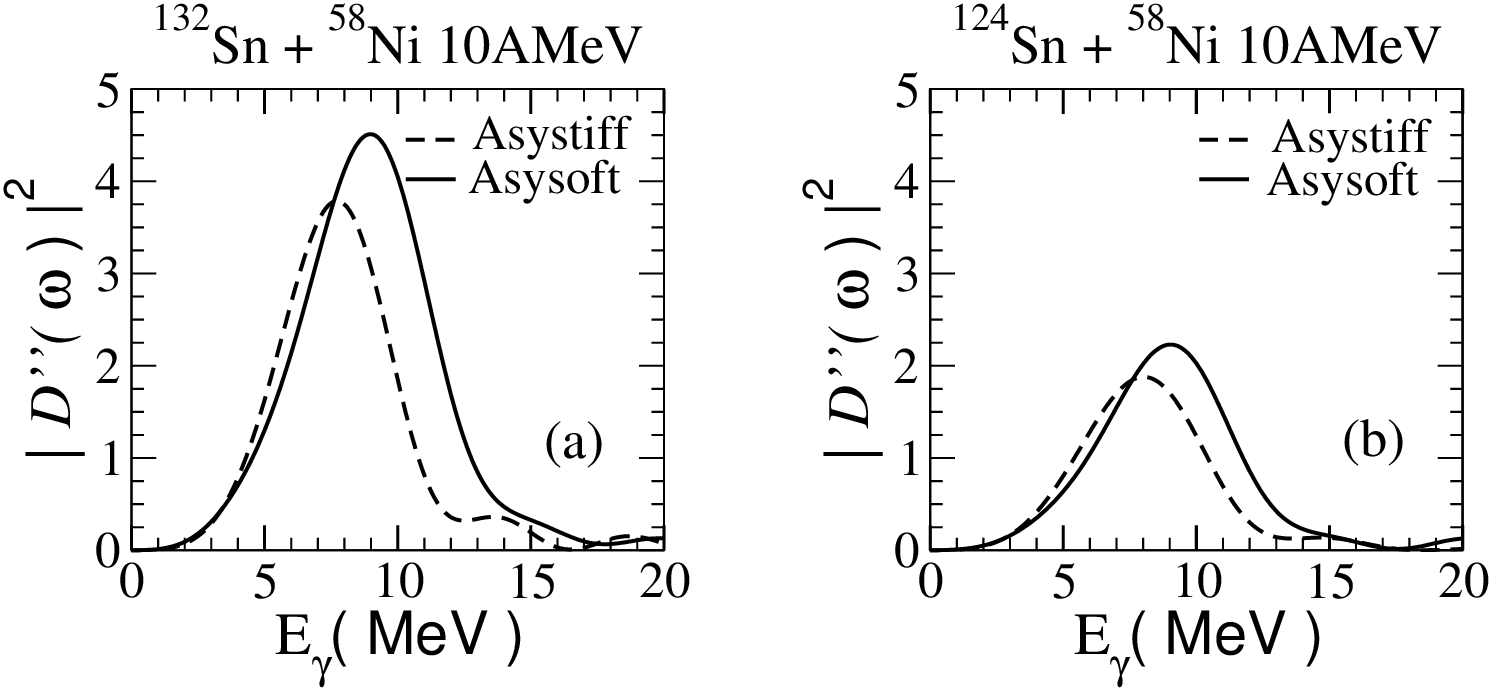}
\end{center}
\caption{(a) panel: Exotic ``132'' system. Power spectra of the 
dipole acceleration at  $b=4$fm (in $c^2$ units).
(b) panel: Corresponding results for the stable ``124'' system.
Solid lines correspond to Asysoft EoS, the dashed to Asystiff EoS.}
\label{yield1}
\end{figure}

In refs. \cite{umar85,umar07} the pre-compound dipole radiation yield is
evaluated via a stationary oscillatory model for a time-dependent charged 
source, described in a TDHF approach. We note that the ``bremss'' procedure
used here is more realistic in the general case of non-equilibrium reaction
dynamics:
in fact while for stationary conditions, when 
$\mid D''(\omega)\mid^2 = \omega^2 \mid D(\omega)\mid^2$, we recover
the same results, see \cite{bar01}, 
in presence of damping effects the transient nature of such fast collective 
mode is consistently accounted for \cite{phonons}. As we will clearly see 
in the following, the damping dynamics, underestimated in a TDHF scheme,
is in fact important for the excitation function of the fast dipole radiation
and for the expected angular anisotropy.

In Fig. 2 (a) we report the power spectrum, 
$\mid D''(\omega) \mid^2$ in semicentral
``132'' reactions, for the different Iso-EoS choices.
The gamma multiplicity is simply related to it, see Eq.(\ref{emission}).
We clearly observe a  
lower value of the centroid,  as well as a reduced total yield, 
in the Asystiff case, due  to the weaker restoring force 
for the dynamical dipole
in the dilute ``neck'' region, where the symmetry energy is smaller
\cite{bar05a}.
Slightly wider distributions are obtained in the Asysoft case, due to the
larger neutron evaporation, that damps the collective oscillation. 
The corresponding results for the stable ``124''  system are drawn
in the (b) panel.
As expected from the larger initial charge asymmetry, 
the Prompt Dipole Emission is increased for the exotic
n-rich beam.

From Eq.(2) we can get the total, energy and impact parameter integrated, 
yield for the two systems and the two Iso-EoS. We find $3.0~10^{-3}$ 
($2.5~10^{-3}$) for $^{124}Sn$ and  $5.7~10^{-3}$($4.4~10^{-3}$) for
 $^{132}Sn$ in the Asysoft (Asystiff) case. 

A detailed analysis of the sensitivity of the results to the symmetry energy
choice 
can be performed just fitting the dipole oscillations by a simple damped 
oscillator 
model, $D(t)=D(t_0) e^{i (\omega_0+i/\tau) t}$, where
$D(t_0)$ is the value at the onset of the collective dinuclear response, 
$\omega_0$ the frequency, that depends on the symmetry energy choice, 
and $\tau$ the damping rate, related to
two-body N-N collisions and neutron emission.
The power spectrum of 
the dipole acceleration is given by 

\begin{equation}
\mid D''(\omega)\mid ^2 =\frac{(\omega_0^2+{1/\tau}^2)^2 {D(t_0)}^2}
 {(\omega-\omega_0)^2 + {1/\tau}^2}
\label{power}
\end{equation}

which from Eq.(\ref{emission}) leads to a total yield proportional to
%\begin{equation}
%\int_{-\infty}^{+\infty} \frac{\mid D''(\omega)\mid ^2}{\omega} d{\omega} =
$\omega_0 \tau (\omega_0^2+{1/\tau}^2){D(t_0)}^2 \simeq 
\omega_0^3 \tau {D(t_0)}^2$
%\label{total}
%\end{equation}
since $\omega_0 \tau >1$.
We clearly see the effect of the Iso-EoS on the total yield, through the
quantity   $\omega_0^3 \tau$, that is slightly dependent on the system.  
Hence, from the above relation, 
the difference of the yields associated with two different 
systems, % ($^{124}Sn$ and $^{132}Sn$),  
that is the quantity usually exploited in the experimental 
analysis \cite{pie05}, 
depends on the Iso-EoS and the sensitivity is amplified when
using  exotic, more asymmetric beams, due to the factor $D(t_0)^2$, 
allowing for a 
clear experimental observation.
It is worthwhile to mention that, according to our fit, we find that
that the parameter $D(t_0)$ may be less than the touching point dipole
amplitude, especially in the Asystiff case and for the exotic
neutron-rich system. 
A delay in the onset of the collective response is expected and so
a more reduced $D(t_0)$ 
with respect to the initial ``geometrical'' value, see also the following.
In fact, we find that the ratio of the total, impact parameter 
integrated yields obtained with the two Iso-EoS (Asysoft relative
to Asystiff) 
is larger in the $^{132}Sn$ case. 
We obtain $1.2$ for the $^{124}Sn$  reaction and $1.3$ in the  
$^{132}Sn$ case.  
This result points to other 
interesting Iso-EoS studies that can be performed
from an accurate measurement of spectrum and yield of the prompt dipole
radiation.

\vskip 0.5cm
\noindent
{\it Anisotropy}

Aside the total gamma spectrum the corresponding
angular distribution can be a sensitive probe to explore the 
properties of preequilibrium dipole mode and the early stages of
fusion dynamics. In fact a clear anisotropy vs. the beam axis
has been recently observed \cite{martin08}.
For a dipole oscillation just along the beam axis we expect an angular 
distribution of the emitted photons like $W(\theta)\sim \sin^2 \theta 
\sim 1+a_2P_2(cos \theta)$ with $a_2=-1$, where $\theta$ is the polar angle
between the photon direction and the beam axis. Such extreme anisotropy
will be never observed since in the collision the prompt dipole axis will 
rotate during the radiative emission. In fact the deviation from the 
$\sin^2 \theta$ behavior will give a measure of the time interval of the fast 
dipole emission.
Just for comparison with statistical compound nucleus GDR
radiation we remind that in the case of a prolate nucleus with a collective 
rotation, for the low energy component we can have an anisotropy parameter 
$a_2=-1/4$, averaging over all 
possible rotation angles and all possible orientations of the collective 
angular momentum (orthogonal to the beam axis), \cite{grbook}.
Orientation fluctuations can even reduce such anisotropy, \cite{alhas90}.

These results cannot be translated 
directly to the case of the dynamical dipole.
As we see from our calculations (Fig. 1) the preequilibrium oscillations 
extend over the first
$250-300 fm/c$. During this time interval, depending also on the 
centrality and energy, 
the deformed nucleus may not complete a full rotation on the reaction plane.
Let us denote by $\phi_i$ and $\phi_f$ the initial and final angles of the
symmetry axis (which is also oscillation axis) with respect to the beam axis,
associated respectively to excitation and complete damping of the dipole mode.
Then  $\Delta \phi=\phi_f-\phi_i$ is the rotation angle 
during the collective oscillations. We can get the angular distribution in 
this case
 by averaging only over the angle $\Delta \phi$ obtaining

\begin{equation}
W(\theta) \sim 1-(\frac{1}{4}+\frac{3}{4}x)P_2(cos \theta)
\label{angdis}
\end{equation}
where $x=cos (\phi_f$+$\phi_i)\frac{sin (\phi_f-\phi_i)}{\phi_f-\phi_i}$ .    

It is easy to see that for  $\Delta \phi = 0$  and $\Delta \phi = 2 \pi$
we recover the two cases discussed above. 
Moreover if $ \phi_f = \phi_i = \phi_0 $ (i.e. the orientation is frozen
at an angle $\phi_0$ with respect to the beam axis)
Eq.(\ref{angdis}) gives an $a_2=-(1-\frac{3}{2} sin^2 \phi_0 )$,
with a change of sign for $\phi_0 \geq 55^\circ$, i.e. a decrease of
$W(\theta)$ around $\theta \approx \pi/2$.
The point is that meanwhile the emission is damped.  

Within the bremsstrahlung approach we can perform an accurate evaluation 
of the prompt dipole angular distribution using a weighted form where the 
time variation of the radiation emission probability is accounted for
\begin{equation}
W(\theta)=\sum_{i=1}^{t_{max}} \beta_i W(\theta,\Phi_i)
\label{wweighted}
\end{equation}
We divide the dipole emission time in $\Delta t_i$ intervals with the 
corresponding $\Phi_i$ mean rotation angles and the related radiation 
emission probabilities
$\beta_i=P(t_i)-P(t_{i-1})$, where 
%\begin{equation}
$P(t)=\int_{t_0}^{t} \mid D''(t) \mid^2 dt / P_{tot}$
%\label{prob}
%\end{equation}
with $P_{tot}$ given by $P(t_{max})$, total emission probability at the
final dynamical dipole damped time.

In Fig.3, (a) panel, we plot the time dependence of the rotation angle,
for the ``132'' system,  
extracted 
from dynamical simulations at $b=2fm$ and $b=4fm$. We note that essentially 
the same
curves are obtained with the two Iso-EoS choices: the overall rotation 
is mostly ruled by
the dominant isoscalar interactions.  
Symmetry energy effects will be induced by the different time evolution 
of emission probabilities. 

\begin{figure}
\begin{center}
\includegraphics*[scale=0.33]{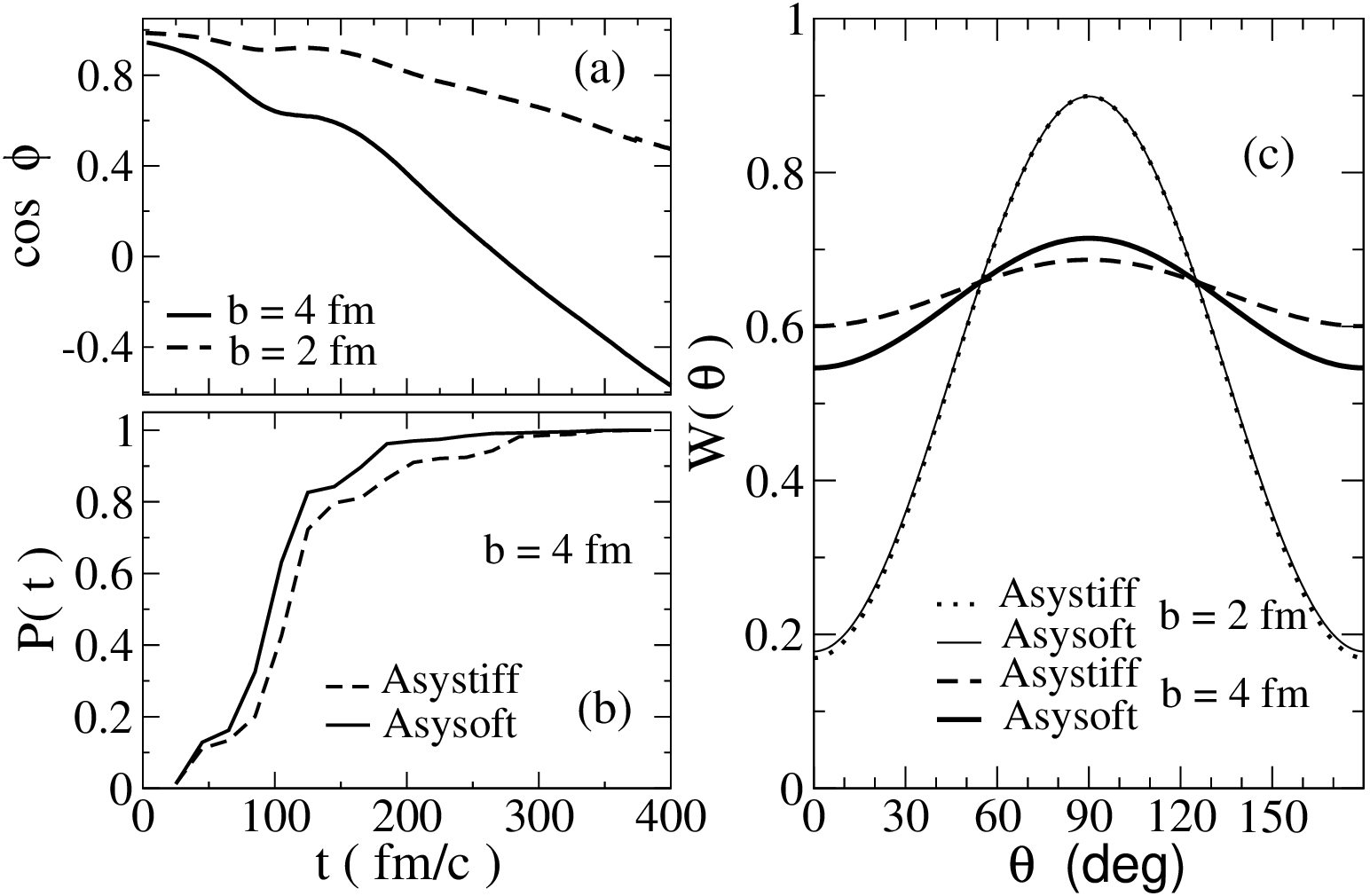}
\end{center}
\caption{ ``132'' system. (a) panel: time dependence of the 
rotation angle at b=2fm 
(dashed line) and b=4fm (solid line).
(b) panel: time evolution of the emission probability 
$P(t)$, see text, for $b=4fm$
impact parameter.
(c) panel: weighted angular distributions for
$b=2fm$ and $b=4fm$ centralities for different symmetry term choices.
Dashed lines for the Asystiff choice, solid for Asysoft.
The Iso-Eos effects on the rotation angle are negligible.
}
\label{yield}
\end{figure}

This is shown in
Fig.3, (b) panel,  
 for the
``132'' system at $b=4fm$ impact parameter. We clearly see that the dominant 
emission region is the initial one, between 50 and 150 fm/c, while the 
dinuclear system rotates of about 20 degrees, roughly from $40^\circ$ 
to $60^\circ$.
Another interesting point is the dependence on the symmetry energy. 
With a weaker symmetry term (Asystiff case) the $P(t)$ is
a little delayed and presents a smoother behavior. As a consequence we can 
expect possible symmetry energy effects even on the angular distributions.

This is shown in the (c) panel of Fig.3, where we have the 
weighted distributions,
Eq.(\ref{wweighted}), for $b=2fm$ and $b=4fm$ impact parameters,
 with the two choices of the symmetry 
energies below saturation. For more central collisions, due to the small
rotation of the oscillation axis, the delay effect in the asystiff case is 
not affecting the angular distribution. For more peripheral reactions
we see a larger contribution at forward/backward angles, although the bump
around $\pi/2$ is still present due to the decreasing emission probability
at later times when the larger rotations contribute. Altogether we get 
wider ``dipole'' angular distributions with respect to the beam axis,
in agreement with the first available data, \cite{martin08}. 
Moreover, as evidenced by the results at b = 4 fm, 
we expect to see a sensitivity to the slope of the symmetry term 
below saturation in presence of larger rotation velocities, i.e. in fusion 
events with
high spin selection.

Summarizing we have shown, within a mean field transport approach,
 that in fusion with exotic nuclei an enhanced 
preequilibrium dipole emission can be observed with a peculiar
angular distribution related to its early emission. The features of this
collective mode are sensitive to the density dependence of symmetry energy  
below saturation. The larger emission with the exotic beam will enhance
the observation of Iso-EoS effects.

The angular distributions are also sensitive to the
fusion dynamics and dipole excitation mechanism and lifetime.
The results presented here can be important to plan new experiments.

In conclusion the dynamical dipole mode appears to be a suitable probe to
test the symmetry energy term in the nuclear EoS as well as to scrutinize
the early entrance channel dynamics in dissipative reactions with radioactive
beams.

{\bf Aknowledgements}

The authors are grateful to D. Santonocito and the Medea Collaboration
for stimulating
discussions. One of authors, V. B. thanks for warm hospitality at Laboratori
Nazionali del Sud, INFN. This work was supported in part by the Romanian
Ministery for Education and Research under the contracts PNII, No.
ID-946/2007 and ID-1038/2008.


\begin{thebibliography}{00}

\bibitem{bao01}
Isospin Physics in Heavy Heavy Ion Collisions at Intermediate Energies,
Eds. Bao-An Li and W. Udo Schroder, Nova Science Publishers, Inc, New York, 
2001).

\bibitem{ste05}
A.W. Steiner,M. Prakash,J.M. Lattimer, P.J. Ellis,
Phys. Rep. 411 (2005) 325.

\bibitem{bar05a}
V. Baran, M. Colonna, V.Greco, M. Di Toro, 
Phys. Rep. 410 (2005) 335.

\bibitem{bao08}
Bao-An Li, Lie-Wen Chen, Che Ming Ko
Phys. Rep. 464 (2008) 113. 

\bibitem{xu00}
H.S. Xu et al.,
Phys. Rev. Lett. 85 (2000) 716.

%\bibitem{bar98}
%V. Baran, M. Colonna, M. Di Toro,A.B. Larionov,
%Nucl. Phys. A 632 (1998) 287.

%\bibitem{tsa01}
%B. Tsang et al.,
%Phys. Rev. Lett. 86 (2001) 5023.

\bibitem{tsa04}
M.B. Tsang et al.,
Phys. Rev. Lett. 92 (2004) 062701.

\bibitem{bar04}
V. Baran, M. Colonna, M. Di Toro,
Nucl. Phys. A 730 (2004) 329.

\bibitem{fil05}
E. De Filippo et al., 
Phys. Rev. C 71 (2005) 044602.

%\bibitem{bar05}
%V. Baran et al.
%Phys. Rev. C 72 (2005) 064620.

%\bibitem{hof79}
%H. Hofmann et al., Z. Phys. A 293 (1979) 229.

%\bibitem{hof79}
%L.G. Moretto, J. Sventek, G. Mantzouranis,
%Phys. Rev. Lett. 42 (1979) 563.

%\bibitem{her81}
%E.S. Hernandez et al., Nucl. Phys. A 361 (1981) 483.

%\bibitem{mdt85}
%M. Di Toro, C. Gregoire,
%Z. Phys. A 320 (1985) 321.

%\bibitem{gre87}
%C. Gregoire et al., Phys. Lett. B 186 (1987) 14.

%\bibitem{sur89}
%E. Suraud et al.,
%Nucl. Phys. A 492 (1989) 294


\bibitem{cho93}
P. Chomaz, M. Di Toro, A. Smerzi,
Nucl. Phys. A 563 (1993) 509.

\bibitem{bar96}
V. Baran et al.,
Nucl. Phys. A 600 (1996) 111.

\bibitem{bon81}
P. Bonche, N. Ngo,
Phys. Lett. B 105 (1981) 17.

\bibitem{umar85}
A.S. Umar et al.,
Phys. Rev. C 32 (1985) 172.

\bibitem{sim01}
C. Simenel, P. Chomaz, G.de France,
Phys. Rev.Lett. 86 (2001) 2971; 
Phys. Rev. C 76 (2007) 024609.

\bibitem{umar07}
A.S. Umar and V.E. Oberacker,
Phys. Rev. C 76 (2007) 014614.

\bibitem{pap03}
M. Papa et al.,
Phys. Rev. C 68 (2003) 034606;
Phys. Rev. C 72 (2005) 064608.


\bibitem{fli}
S.Flibotte et al., Phys. Rev. Lett. 77 (1996) 1448.

\bibitem{pie05}
D. Pierroutsakou et al., Phys. Rev. C 71 (2005) 054605.


\bibitem{bar01b}
V. Baran et al.,
Nucl. Phys. A 679 (2001) 373.

\bibitem{bar01}
V. Baran, D.M. Brink, M. Colonna, M. Di Toro,
Phys. Rev. Lett. 87 (2001) 182501.


%\bibitem{col98}
%M. Colonna et al., Phys. Rev. C 57 (1998) 1410.

%\bibitem{bar02}
%V. Baran et al., Nucl. Phys. A 703 (2002) 603.

\bibitem{li93}
G.Q. Li, R. Machleidt,  Phys. Rev. C 48 (1993) 1702;
 Phys. Rev. C 49 (1994) 566.

\bibitem{class08}
The classical "bremss" emission probability Eq.(2) appears justified by the 
fact
that in this dipole mode nucleons are oscillating in a classical allowed 
region.
Moreover in equilibrium conditions we can recover the usual statistical 
radiation
probability for a GDR emission from a compound nucleus at a given temperature,
see \cite{bar01}.

\bibitem{phonons}
A decaying phonon model was worked out in refs.\cite{cho93,bar96}. A 
comparison with the ``bremss'' approach can be found in ref.\cite{bar01}. 
Some unconsistency appears: the larger
``bremss'' yields can be reproduced only adjusting the number of GDR phonons 
to values corresponding to times well before statistical equilibration.

%\bibitem{rizzoj08}
%J.Rizzo et al., Nucl. Phys. A 806 (2008) 79-104.

\bibitem{martin08}
B. Martin, D. Pierroutsakou et al. (Medea Collab.),
Phys. Lett. B 664 (2008) 47

\bibitem{grbook}
M.N. Harakeh, A. van der Woude, {\it Giant Resonances}, 
 Oxford Univ.Press 2001

\bibitem{alhas90}
Y. Alhassid, B. Bush, Phys. Rev. Lett. 65 (1990) 2527




\end{thebibliography}
\end{document}